\def\year{2023}\relax
\documentclass[letterpaper]{article} 
\usepackage{aaai22}  
\usepackage{times}  
\usepackage{helvet}  
\usepackage{courier}  
\usepackage[hyphens]{url}  
\usepackage{graphicx} 
\usepackage{natbib}  
\usepackage{caption} 
\DeclareCaptionStyle{ruled}{labelfont=normalfont,labelsep=colon,strut=off} 
\usepackage{algorithm}
\usepackage{algorithmic}
\usepackage{newfloat}
\usepackage{listings}
\floatstyle{ruled}
\newfloat{listing}{tb}{lst}{}
\floatname{listing}{Listing}

\usepackage{xcolor}
\definecolor{snp}{HTML}{FDF38E}
\definecolor{sgp}{HTML}{00B140}
\definecolor{con}{HTML}{0087DC}
\definecolor{lab}{HTML}{E4003B}
\definecolor{lib}{HTML}{FAA61A}
\definecolor{pc}{HTML}{005B54}
\definecolor{sf}{HTML}{326760}
\definecolor{dup}{HTML}{D46A4C}
\definecolor{sdlp}{HTML}{2AA82C}
\definecolor{apni}{HTML}{F6CB2F}
\definecolor{tuv}{HTML}{0C3A6A}
\definecolor{pbp}{HTML}{E91D50}

\usepackage{multirow}
\usepackage{tabularx}
\usepackage{adjustbox}
\usepackage{subcaption}
\usepackage{booktabs}

\usepackage{comment}

\usepackage{color}

\title{Generalizing Political Leaning Inference to Multi-Party Systems:\\Insights from the UK Political Landscape}

\author {
    Joseba Fernandez de Landa,\textsuperscript{\rm 1}
    Arkaitz Zubiaga, \textsuperscript{\rm 2}
    Rodrigo Agerri \textsuperscript{\rm 1}
}
\affiliations {
    \textsuperscript{\rm 1} HiTZ Center - Ixa, University of the Basque Country UPV/EHU\\
    \textsuperscript{\rm 2} Queen Mary University of London\\
    joseba.fernandezdelanda@ehu.eus, a.zubiaga@qmul.ac.uk, rodrigo.agerri@ehu.eus 
}

\usepackage{bibentry}

\begin{document}

\maketitle

\begin{abstract}
An ability to infer the political leaning of social media users can help in gathering opinion polls thereby leading to a better understanding of public opinion. While there has been a body of research attempting to infer the political leaning of social media users, this has been typically simplified as a binary classification problem (e.g. left vs right) and has been limited to a single location, leading to a dearth of investigation into more complex, multiclass classification and its generalizability to different locations, particularly those with multi-party systems. Our work performs the first such effort by studying political leaning inference in three of the UK's nations (Scotland, Wales and Northern Ireland), each of which has a different political landscape composed of multiple parties. To do so, we collect and release a dataset comprising users labelled by their political leaning as well as interactions with one another. We investigate the ability to predict the political leaning of users by leveraging these interactions in challenging scenarios such as few-shot learning, where training data is scarce, as well as assessing the applicability to users with different levels of political engagement. We show that interactions in the form of retweets between users can be a very powerful feature to enable political leaning inference, leading to consistent and robust results across different regions with multi-party systems. However, we also see that there is room for improvement in predicting the political leaning of users who are less engaged in politics.
\end{abstract}

\section{Introduction}
Ideology is a set of people's beliefs that can be understood as ways of thinking and acting in society. Those beliefs can generally be represented by political parties, acting like social hubs of coordinated thoughts and actions. However, ideology is often presented and simplified into binary frameworks based on individuals stance over left/right or conservative/liberal orientations. Political leaning inference is proposed as a way of representing individual actors by the closest political party, analyzing ideology from a richer perspective. To better understand society, social researchers can benefit from the development of efficient methods capable of generalizing political leaning inference across different regions \cite{imhoff2022conspiracy}. We address this challenge by investigating new tools and techniques for conducting deeper and more accurate social and political research with the aim of improving opinion polls, political polarization studies, stance and propaganda detection or disinformation analysis among others.

The vast majority of research on political leaning inference has been limited to binary classification between the two prevailing parties or stances \citep{conover2011political, conover2011predicting,Barber2015UnderstandingTP,Barber2015BirdsOT,garimella2017long,barbera2015tweeting,Pennacchiotti2011DemocratsRA, Hua2020TowardsMA, timme20}. The few works that conducted multiclass classification \citep{Boutet2012WhatsIY,Makazhanov2013PredictingPP,Rashed2021EmbeddingsBasedCF} were constrained to a single scenario or region. This however limits both the applicability of such methods and the insights learned from such studies. Indeed, each social context has its own political reality which is typically reflected in more than two ideological options. 

In order to broaden the study on the ability to infer the political leaning of social media users, we identify four key limitations in previous work. First, the widely used binary frameworks can be limiting and imprecise as individuals hold a wider range of political beliefs, which instead calls for multiclass classification. Second, political ideologies and beliefs can vary substantially across different regions since each community has its own idiosyncrasy, which calls for the study of applicability across regions. Third, to achieve generalizability, it is crucial to include scenarios where the labeled data available for training is limited, which makes critical the study of few-shot learning approaches. Fourth, it is important to consider that not every social media user is as engaged in politics and/or is vocal about their beliefs, which posits the importance of assessing the ability to predict the political leaning of all kinds of users regardless of their level of engagement.

By addressing the above limitations, we aim to further research in political leaning inference by providing the first study that focuses on generalizing a multiclass political leaning inference model across different scenarios or regions. With this goal in mind, we propose and evaluate a range of techniques for data extraction and user representation for multiclass political leaning inference. The aim is to independently represent Twitter users by leveraging their interactions, effectively transforming content sharing actions on Twitter into vector spaces. By accomplishing this, we seek to achieve adaptability across diverse situations, in turn opening an avenue for further exploration in other tasks. Thus, retweet interactions are selected to represent users, known for their effectiveness in achieving user classification \citep{conover2011predicting, magdy16, darwish2020unsupervised, Stefanov2020PredictingTT, fernandez2022relational}. We conducted experiments using four distinct unsupervised techniques, namely ForceAtlas2 \citep{jacomy2014forceatlas2}, DeepWalk \citep{deepwalk}, Node2vec \citep{Grover2016node2vecSF} and Relational Embeddings \citep{fernandez2022relational}. Those user representations are evaluated in three different regions, each with different political parties, including also the first study on the ability of those techniques to rely on scarce training data as well as to determine the political leaning of users with different levels of engagement.

To the best of our knowledge, our work is the first to study the political leaning inference task across diverse multiclass political realities, which in turn leads to new insights into tackling this challenging setting. More specifically, this paper makes the following novel contributions:

\begin{itemize}
 \item we devise and experiment with a pluralistic framework that includes multiple political leanings, which proves adaptable to different regions since it is grounded on localized political actors;
 \item we propose and evaluate a range of methodologies to make the most of retweet interactions among social media users to infer their political leaning, showing that Relational Embedding based approach is effective even in weakly-supervised and realistic scenarios; 
 \item we perform a comprehensive error analysis and feature visualizations in order to show the ability of the proposed methodology to capture socio-political information coherently and in alignment with the specific political context.
 \item data resources such as labeled and interaction-based data and code will be made publicly available to facilitate reproducibility of results and enable further research.
\end{itemize}

Through experiments on three of the United Kingdom's regions (Wales, Scotland and Northern Ireland), we find that the use of interaction-pairs leads to competitive performance on political leaning inference when compared to random-walk based approaches. In addition to improved performance, our analysis shows that the resulting visualizations and errors obtained from this approach can provide valuable insights into the political context in the UK.

\section{Related Work} \label{sec:rel_work}
In this section we discuss prior work on political leaning inference and stance detection across different countries, and different methods to extract features from interaction-based data such as retweets between users.

\subsection{Political leaning inference}
\label{sec:rel_work_pol_lean}

Political leaning of elected representatives has been approached using roll call votes on parliaments of bipartite systems such as US \cite{akoglu2014quantifying} or even multy-party systems like Brazil \cite{10.1371/journal.pone.0140217}. Whereas elected people's political leaning can be inferred based on their public votes, the extension of these studies is limited to the specific parliaments and can not be extended to other populations as can be done with social media data. 

Political leaning inference from social media was pioneered by \citet{conover2011political}, who framed it as a left-right dichotomy in the US context by using retweets. Subsequently, others have followed a similar categorization of users into left or right, such as \citet{Barber2015UnderstandingTP} and \citet{Barber2015BirdsOT} in the context of elections, \citet{barbera2015tweeting} classifying users as liberal or conservative, or \citet{garimella2017long} focused on political accounts and media outlets. In addition, Twitter users from the USA have been classified as Democrats or Republicans \citep{Pennacchiotti2011DemocratsRA, Hua2020TowardsMA} and, although some other work has gone beyond the strictly binary classification by proposing a seven-point scale to place users in a conservative-liberal spectrum \citep{PreotiucPietro2017BeyondBL}, they still adhere to the same dichotomy.

Multiclass classification in more diverse political landscapes, the most common situation for a large number of countries, has barely been studied. This limits the ability to evaluate existing methods in those scenarios as well as the capacity to learn new insights from social media for those specific contexts.
  
Political ideology can also be approached as, or associated with, the arguably more popular stance detection task \citep{mohammad-etal-2016-semeval, hardalov-etal-2021-cross}. Still, much of the stance detection research has focused on topics relevant to politics but not explicitly on political leaning inference. This is the case, for example, of recent shared tasks and datasets such as SardiStance \citep{cignarella2020sardistance} and VaxxStance \citep{vaxxstance2021}, which proposed stance detection tasks associated with the sardines social movement and the stance towards vaccination. The datasets released by these shared tasks allowed researchers to leverage not only textual content but also social interactions \citep{TextWiller,QMUL-SDS,WordUp,fernandez2022relational}, highlighting the importance of interactions in determining the stance of users who form homophilic connections with one another.

Stance detection in social media has also been addressed as an unsupervised task by relying on interactions between users. Previous works following this approach have applied a force-directed algorithm \citep{Fruchterman1991GraphDB} or other methods such as UMAP \citep{McInnes2018UMAPUM}. For example, \citet{darwish2020unsupervised} used both force-directed algorithm and UMAP for unsupervised stance detection on Twitter users using retweets. Furthermore, UMAP has also been used to get interaction-based retweet features for automatically tagging Twitter users' stance on different topics \citep{Stefanov2020PredictingTT} and to study political polarization in Turkey \citep{Rashed2021EmbeddingsBasedCF}. A shortcoming of these approaches is that, in order to be able to handle the huge interaction networks, the features are based only on a set of users manually picked as being more salient.

\subsection{Approaches to modeling user interactions}

User interactions in social media have been used to study political disinformation in the 2018 presidential election in Brazil, based on opposed hashtags \citep{soares2021hashtag}, or to identify the roles that users play in political conversations during polarized online discussions \citep{recuero2019using}. Besides, interaction features have also been useful to map Persian Twitter during Iran's 2017 presidential election by investigating the network structures generated by the users and their sharing practices \citep{kermani2021mapping}. Finally, \citet{Zubiaga2019PoliticalHI} studied stance detection to analyze independence movements in Catalonia, Basque Country and Scotland, showing that features extracted from the followers network obtained the best performance. Following the same idea, an analysis of young Basque Twitter users were also mapped based on their retweet-based data \citep{basqueyoungtwitter}. 

Neural approaches for learning node representations on user interaction data include DeepWalk \citep{deepwalk} and node2vec \citep{Grover2016node2vecSF} among the most popular and effective choices \citep{socialphysics2022,9565320}. These node-representation features are created based on unlabeled data, characterizing the nodes as low-dimensional features. These methods try to predict a group of neighboring users that emerge from Random Walks, based on the input of a given user. This means that they focus more on the overall structure of the interaction network instead of trying to model pair-based interactions. An issue to consider is that random walks might create artificial interactions that may not actually occur in the interaction pairs. An alternative proposal to build interaction-based models is the Relational Embeddings method, based on real interaction-pairs applied to predict users' stance \citep{fernandez2022relational}.

Finally, other neural approaches such as graph convolutional networks (GCN) \citep{Kipf2017SemiSupervisedCW} and graph attention networks (GAT) \citep{Velickovic2018GraphAN} require previously obtained feature representations to train end-to-end models. Therefore, these neural models require labeled data and additional features during the feature learning process. As a result, these models cannot be effectively employed to extract user representations using an unsupervised approach and are left out of our study. TIMME is a technique developed for identifying Democrat/Republican leaning on Twitter by utilizing multi-task learning and multi-relational data such as follow, retweet, reply, mention and like \citep{timme20}. However, testing these three algorithms was not possible with the hardware we currently have at our disposal (see Table \ref{tab:data_final} for size of our datasets).

\section{The Political Context in the UK}\label{sec:regsel}

Given our interest in analyzing the socio-political context of the United Kingdom as a multi-party system \citep{lynch2007party}, our study focuses on political parties in Scotland (5.5M citizens), Wales (3.1M) and Northern Ireland (1.8M). These regions form politically diverse contexts, each with its own devolved government and strong nationalist sentiments that foster many political options. The UK's political landscape has evolved substantially in recent decades, from being dominated by two parties in the 1950s (Conservative and Labour parties attaining over 95\% of the votes) to a more diverse, multi-party landscape (75\% across both parties in 2019).

\textbf{Scotland (SCT):} \textit{Scottish National Party} (SNP \textcolor{snp}{$\bullet$}) is a Scottish nationalist and social democratic political party; positioned on the center-left, pro-independence and pro-European. \textit{Scottish Conservative \& Unionist Party} (SCU \textcolor{con}{$\bullet$}) is a conservative party in Scotland, Nationally affiliated with the Conservative Party; centre-right and unionist. \textit{Scottish Labour Party} (SL \textcolor{lab}{$\bullet$}) is a Scottish social democratic political party, an autonomous section of the UK Labour Party; considered to be centre-left and unionist. \textit{Scottish Green Party} (SGP \textcolor{sgp}{$\bullet$}) is a Scottish green political party, affiliated with the Global Greens and associated mainly with environmentalist policies; positioned on the left, pro-independence and pro-European. \textit{Scottish Liberal Democrats} (SLD \textcolor{lib}{$\bullet$}) is a Scottish liberal and federalist political party, part of the United Kingdom Liberal Democrats; positioned on the political centre, pro-European and unionist.

\textbf{Wales (WAL):} \textit{Welsh Labour} (WL \textcolor{lab}{$\bullet$}) is a Welsh social democratic political party, and formally part of the UK Labour Party; centre-left and unionist. \textit{Welsh Conservatives} (WC \textcolor{con}{$\bullet$}) is a conservative party in Wales, a branch of the UK's Conservative Party; ideology is centre-right and unionist. \textit{Plaid Cymru} (PC \textcolor{pc}{$\bullet$}) is the principal Welsh nationalist political party; positioned on the left and pro-independence. \textit{Welsh Liberal Democrats} (WLD \textcolor{lib}{$\bullet$}) is a Welsh liberal and federalist political party, branch of the UK's Liberal Democrats; positioned on the political centre, pro-European and unionist.

\textbf{Northern Ireland (NIR):} \textit{Sinn Féin} (SF \textcolor{sf}{$\bullet$}) is an Irish republican and democratic socialist political party; considered to be left-wing, pro-unification and pro-independence. \textit{Democratic Unionist Party} (DUP \textcolor{dup}{$\bullet$}) is a conservative and loyalist political party in Northern Ireland; positioned on the right-wing and unionist. \textit{Alliance Party of Northern Ireland} (APNI \textcolor{apni}{$\bullet$}) is a liberal political party in Northern Ireland, aligned with the UK's Liberal Democrats; positioned on the political centre, pro-European and they consider themselves outside of Nationalism and Unionism. \textit{Ulster Unionist Party} (UUP \textcolor{con}{$\bullet$}) is a unionist political party in Northern Ireland, a branch of the UK's Conservative Party; positioned on the center-right and unionist. \textit{Social Democratic and Labour Party} (SDLP \textcolor{sdlp}{$\bullet$}) is a social-democratic and Irish nationalist political party in Northern Ireland; centre-left and pro-Irish.

\section{Datasets}\label{sec:datacol}

We devise a generalizable data collection methodology, in our case tested in the UK but extensible to other regions. Once the regions of interest and the relevant political parties have been identified, our methodology consists of three steps to collect: (i) an initial seed of users (members), (ii) other users with different levels of engagement or interest (supporters and sympathizers) and, (iii) interactions and timelines pertaining to those users. Data collection was done between September and October 2022.

\paragraph{\textbf{Step 1. Manual labeling of seed users.}} In line with data collection strategies followed in previous work \citep{Makazhanov2013PredictingPP, Barber2015BirdsOT, garimella2017long, Hua2020TowardsMA}, we start by collecting an initial seed of users. For each of the political parties in our datasets, we identify party members with Twitter accounts including members of parliament (MPs) or members of regional parlaments (MSPs, MSs and MLAs). This leads to a collection of users where each user is linked to a specific region and party (details in Table \ref{tab:labels}, column `Members').

\begin{figure}[ht]
\centering
    \includegraphics[width=1\linewidth]{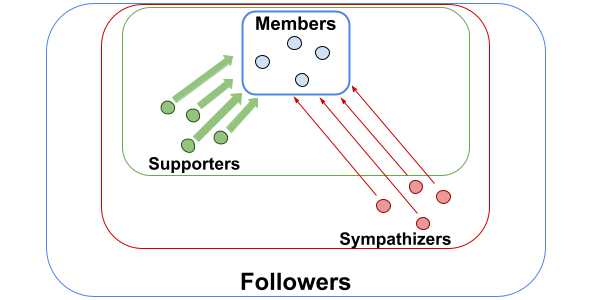}
    \caption{Creation scheme for Supporter and Sympathizer evaluation sets. 
    Supporters: more engaged, following 5 or more member users.
    Sympathizers: less involved, following up to 2 member users.}
    \label{fig:followers}
\end{figure}

\paragraph{\textbf{Step 2. Snowball collection of friends and followers.}} To expand the datasets beyond direct members of the party, we look for less engaged users including supporters and sympathizers (see Figure \ref{fig:followers}). We rely on \emph{follower} connections with member users as a proxy to collect less engaged users \citep{Barber2015BirdsOT, timme20}, where the level of engagement or political interest is determined by the number of members they follow \citep{timme20}. Engaged users with a strong interest in politics are referred to as 'supporters' if they follow 5 or more members of a party. On the other hand, users with less vested interest, such as those who follow 2 or fewer members, are called `sympathizers'. The specific thresholds of 2 and 5 were empirically determined by looking at the frequencies in the data so that we could obtain a balanced number of supporter and sympathizer user groups (similarly done in previous work such as \citet{timme20}). We retrieve up to 100 users per party for each of the supporter and sympathizer groups, filtering out those for which interaction data (see step 3) is not available, leading to the counts shown in the columns `Supporter' and `Sympathizer' of Table \ref{tab:labels}.

\begin{table}[ht!]\small
\centering
\begin{tabular}{@{}lrrrr@{}}
\toprule
Region   & Party     &      Member  &    Supporter &   Sympathizer \\ \midrule
\multirow{6}{*}{SCT}     
&SNP \textcolor{snp}{$\bullet$} &184&96&85 \\             
&SCU \textcolor{con}{$\bullet$} &59&97&84  \\ 
&SL \textcolor{lab}{$\bullet$} &52&95&86  \\  
&SGP \textcolor{sgp}{$\bullet$} &42&99&88 \\      
&SLD \textcolor{lib}{$\bullet$} &24&98&94 \\ 
&total &361&485&437 \\ \midrule
\multirow{5}{*}{WAL}     
&WL \textcolor{lab}{$\bullet$}  &55&97&88\\               
&WC \textcolor{con}{$\bullet$}  &42&98&85\\      
&PC \textcolor{pc}{$\bullet$}  &42&99&85\\ 
&WLD \textcolor{lib}{$\bullet$}  &27&100&95\\  
&total &166&394&353\\  \midrule
\multirow{6}{*}{NIR}     
&SF \textcolor{sf}{$\bullet$}      &80&98&63 \\
&DUP \textcolor{dup}{$\bullet$}    &65&75&67 \\
&APNI \textcolor{apni}{$\bullet$}  &52&83&79 \\
&UUP \textcolor{con}{$\bullet$}    &58&73&68 \\
&SDLP \textcolor{sdlp}{$\bullet$}  &59&76&72 \\
&total &314&405&349\\\bottomrule
\end{tabular}
\caption{Manually labeled Member users and automatically labeled Supporter and Sympathizer users, by region and class.}
\label{tab:labels}
\end{table}

\begin{figure}[ht]
\centering
    \includegraphics[width=1\linewidth]{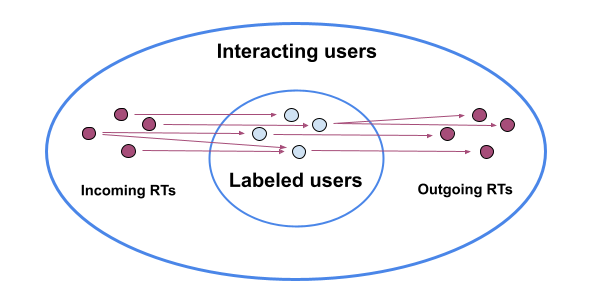}
    \caption{Interacting user's identification scheme. Central circle represents users manually or automatically labeled. External circle illustrates all the users interacting with the labeled users by retweeting them (incoming) or being retweeted by them (outgoing).}
    \label{fig:interacting}
\end{figure}

\paragraph{\textbf{Step 3. Twitter timeline retrieval.}} For every user in the member, supporter and sympathizer groups we retrieve a history of their retweet interactions, regardless of whether these interactions are with users included in our datasets (see Figure \ref{fig:interacting}). The purpose of this retrieval is to identify all users who have interacted with the labeled users through retweets, referred to as \textit{interacting users}. Subsequently, we collected all available retweets from the timelines of both the labeled users and the interacting users, thereby extracting a substantial amount of interactions between pairs of users. Table \ref{tab:data_final} shows the final statistics of retweets retrieved and the number of total users performing those interactions.

\begin{table}[ht!]\small
\centering
\begin{tabular}{@{}lrrr@{}}
\toprule
                    & SCT     & WAL       & NIR   \\ \midrule
Member users        & 361     & 166       & 314   \\ 
Supporter users     & 485     & 394       & 405   \\ 
Sympathizer users   & 437     & 353       & 349   \\ \midrule
Interacting users   & 87k     & 62k       & 21k    \\ 
Retweets            & 19M     & 21M       & 4M  \\ 
All users           & 937k    & 933k      & 426k\\ \bottomrule
\end{tabular}
\caption{Final dataset composition for each region.}
\label{tab:data_final}
\end{table}

\section{Methods}

We experiment with different unsupervised user representation methods in order to represent users based solely on interactions. Those features will be used along with different classification algorithms, to assess their effectiveness in embedding socio-political information. We also evaluate the impact of dimensionality reduction techniques.

\subsection{User Representation Methods}\label{sec:interaction-methods}

We experiment with a set of user representation methods based on leveraging retweet interactions that can represent users' preferences and behavior. Thus, retweet based interactions on their own have been shown to be effective for user classification tasks \citep{conover2011predicting, magdy16, darwish2020unsupervised, Stefanov2020PredictingTT, fernandez2022relational} and have been used to represent users as done in different approaches \citep{cignarella2020sardistance,vaxxstance2021,fernandez2022relational,darwish2020unsupervised}.

Next, we describe the proposed user representation methods which are suited to transform large and heterogeneous data sources.

\textbf{ForceAtlas2} \citep{jacomy2014forceatlas2} is a continuous graph layout algorithm, that transforms a network into a 2-dimensional space to obtain a readable shape. This algorithm orders all the nodes represented in a graph according to the established relations, being a redesign of existing force-directed algorithm \citep{Fruchterman1991GraphDB}. As a result of that approximation-repulsion process, those nodes that are unrelated repulse each other, while related ones will attract each other. 

\textbf{DeepWalk} \citep{deepwalk} algorithm learns user representations by simulating uniform random walks among the connected nodes of the network. Based on Skip-gram \citep{mikolov2013efficient}, being given an instance the context or neighbours has to be predicted. The context for each instance is generated by the random walks among the surrounding connected data points of the instance. The length and number of walks for each instance will determine the context of such instance. Based on local information the method is able to learn representations which encodes structural regularities.

\textbf{Node2vec} \citep{Grover2016node2vecSF} algorithm is similar to DeepWalk, but adds two parameters to control the structure of the network due to a search bias while random walks happen. Those parameters are the return ($p$) and in-out ($q$) parameters. Return parameter ($p$) controls the probability to return to visited points in the random walks, at higher values the probability of revisiting a node decreases. The in-out parameter ($q$) controls the probability to explore undiscovered parts of the graphs, higher values are related to further points.

\textbf{Relational Embeddings} \citep{fernandez2022relational} are based on a single hidden-layer neural network trying to predict who retweeted whom for all the gathered interaction pairs. Instead of generating random walks among nearest neighbours, in this method the interactions are based on the relations between two users. This allows to capture social information and to build meaningful representations based on real user interactions.

The obtained retweet-based interactions are used to feed the aforementioned methods to train the unsupervised models. We use all the available users in order to embed as much information as possible. The user author of the retweet is considered the source of the interaction, while the user receiving the retweet would be the target. We generate source-target user pairs in this manner to provide the input to the user representation methods. With the intention of generating low dimensional but meaningful representations and based on previous work \citep{darwish2020unsupervised,Stefanov2020PredictingTT,fernandez2022relational}, dimensions were set on 20 dims for DeepWalk, Node2vec and Relational Embeddings. The hyperparameters for node2vec and DeepWalk were set to the default values typically used by these algorithms: walks\_per\_node = 10, walk\_length = 80, window or context\_size = 10, and the optimization is executed for a single epoch \citep{deepwalk,Grover2016node2vecSF}. For node2vec, we set p=1 and q=0.5 in order to enhance network community related information \citep{Grover2016node2vecSF}. Default parameters were set for ForceAtlas2 and independent models were trained for each of the regions.

\subsection{Dimensionality Reduction Techniques} \label{sec:dim_red}

We want to study the performance of the best user representation method also for settings in which only limited labeled data is available, namely, in what we refer to as a weakly supervised scenario. We employ various dimension reduction techniques to compress information for weakly supervised settings with the aim of forcing our model to generalize more from the characteristics seen in the training data. We use 3 different dimensionality reduction methods:

\textbf{PCA} is a linear algorithm for dimensionality reduction which aims to represent the data in a low-dimensional space while preserving the original global structure of the data. As this is a linear algorithm, it will only find linear dependencies or relationships in the data, without considering the neighbours on their own.

\textbf{t-SNE} \citep{van2008visualizing} is a nonlinear dimensionality reduction technique that embeds high-dimensional data into a lower dimensional space. A probability distribution is built over data point pairs, giving similar data points a high probability while dissimilar ones are assigned a lower one. The algorithm computes the probability that pairs of data points in the higher dimensional space are related, and then chooses low-dimensional embeddings which produce a similar distribution based on the Kullback–Leibler divergence. 

\textbf{UMAP} \citep{McInnes2018UMAPUM} or Uniform manifold approximation and projection is also a nonlinear dimensionality reduction technique. mThis technique is similar to t-SNE, but it assumes that the data is uniformly distributed on a locally connected Riemannian manifold. UMAP creates a fuzzy graph that reflects the topology of the high dimensional graph based on the nearest neighbours of each data point. Then the low dimensional graph is built based on the fuzzy graph. This dimension reduction technique is able to reflect the large scale global structure, while also preserving the local structure. 

The inputs are user vectors derived from the selected user representation method as presented in the previous subsection. Every dimension reduction techniques are used with their default hyper-parameters. Dimensionality is set to 2, following previous work \citep{darwish2020unsupervised,Stefanov2020PredictingTT} and separate models are trained for each of the regions.

\begin{table*}[htb!]
\small
\centering
\begin{tabular}{@{}l|lllll|lllll|lllll@{}}
\toprule
\multicolumn{1}{c}{}
& \multicolumn{5}{c}{SCT}  & \multicolumn{5}{c}{WAL}  & \multicolumn{5}{c}{NIR} 
\\ \midrule
&Mj&N2V&DW&FA2&RE  &Mj&N2V&DW&FA2&RE &Mj&N2V&DW&FA2&RE 
\\ \midrule
LogReg   
& 13.5 & 71.5  & 68.0  & 31.0  & \textbf{99.4}
& 12.4 & 55.2  & 62.8  & 24.6  & \textbf{99.2}
& 8.1 & 51.5  & 64.8  & 28.7  & \textbf{97.7}
\\
RF    
& 13.5 & 81.8  & 78.2  & 63.5  & \textbf{99.2}
& 12.4 & 67.5  & 78.2  & 68.9  & \textbf{96.2}
& 8.1 & 66.4  & 80.9  & 76.3  & \textbf{97.4} 
\\
NB        
& 13.5 & 37.5  & 35.5  & 51.1  & \textbf{99.5}
& 12.4 & 30.3  & 32.7  & 42.3  & \textbf{98.7}
& 8.1 & 28.0  & 32.1  & 34.0  & \textbf{97.7} 
\\
SVM - linear    
& 13.5 & 73.0  & 69.0  & 30.6  & \textbf{99.4}
& 12.4 & 33.2  & 54.7  & 37.9  & \textbf{96.4}
& 8.1 & 35.4  & 46.9  & 41.2  & \textbf{97.7} 
\\
SVM - poly.     
& 13.5 & 39.7  & 43.2  & 30.0  & \textbf{96.4}
& 12.4 & 22.2  & 25.9  & 12.6  & \textbf{96.4}
& 8.1 & 08.1  & 10.3  & 10.3  & \textbf{94.5} 
\\
SVM - rbf       
& 13.5 & 41.3  & 43.1  & 61.6  & \textbf{99.9}
& 12.4 & 26.1  & 28.6  & 38.4  & \textbf{98.6}
& 8.1 & 17.8  & 29.7  & 50.6  & \textbf{97.4}  
\\ \midrule
average       
& 13.5 & 57.5  & 56.2  & 44.6  & \underline{\textbf{98.9}}
& 12.4 & 39.1  & 47.2  & 37.5  & \underline{\textbf{97.6}}
& 8.1 & 34.5  & 44.1  & 40.2  & \underline{\textbf{97.0}}  
\\ \bottomrule
\end{tabular}
\caption{F1 macro score results leave-one-out CV on SCT, WAL and NIR member datasets. Mj refers to majority label classifier used as baseline. Algorithms used to generate the representations: N2V (Node2vec), DW (DeepWalk), FA2 (ForceAtlas2), RE (Relational Embeddings). Underlined values represent statistically significant ($p$ $<$ $0.05$) highest average values.}
\label{tab:f1_results_sup}
\end{table*}

\section{Experiment \#1: Strongly Supervised Scenario}

\paragraph{\textbf{Experiment Settings.}} For each region, we experiment with the users in the group of members using a leave-one-out (LOO) cross-validation setting, i.e., one user is left for testing while all the others are used for training. Thus, the model is trained and tested once for each user in the dataset, which is feasible given the low dimensionality of the representations. The primary objective of this experiment is to compare the performance of user representation methods. The user representations obtained using the different methods are used to train six classification algorithms: Logistic Regression (LogReg), Random Forest (RF), Naive Bayes (NB) and linear, polynomial and RBF-kernel Support Vector Machines (SVM). We use their scikit-learn implementation \citep{scikit-learn} with default configuration.

\paragraph{\textbf{Results.}} Looking at the results reported in Table \ref{tab:f1_results_sup}, we observe that the models trained with RE representations achieve best results for all the classifiers across every region. Among the models trained with RE representations, Logistic Regression consistently obtains the best results. On the other hand, and despite its popularity, the FA2 representations lead to the lowest performance scores, showing that it is the least suitable for this task. Both N2V and DW are clearly better than FA2, but still are clearly outperformed by the models obtained with RE representations. The RE method also behaves more robustly in multi-party scenarios and across regions. Finally, N2V, DW and FA2 show substantial variability across the different regions, while RE is the most stable method, showing robustness and adaptability.

\begin{table}[h!]\small
\centering
\begin{tabular}{@{}lrrrr@{}}
\toprule
     & Dim. Red.     &      LOO  &    3-shot &   1-shot \\ \midrule
\multirow{4}{*}{SCT}     
& none   &        99.4   &         71.8     &        91.9 \\
     & UMAP  2d  &\textbf{99.9}  &         91.2     &        90.8  \\
     & t-SNE 2d  &        99.4   & \textbf{95.3}    &\textbf{92.2} \\ 
     & PCA   2d  &        87.3   &         71.2     &        67.4 \\\midrule
\multirow{4}{*}{WAL}     
& none   &\textbf{99.2}  &         96.9     &\textbf{98.5}  \\
     & UMAP  2d  &        98.5   & \textbf{97.8}    &        94.0 \\
     & t-SNE 2d  &        98.5   & \textbf{97.7}    &        96.2 \\ 
     & PCA   2d  &        93.9   &         89.5     &        75.3 \\\midrule
\multirow{4}{*}{NIR}     
& none   &\textbf{97.7}  &         97.0     &        94.6 \\
     & UMAP  2d  &        97.3   & \textbf{97.2}    &        94.7  \\
     & t-SNE 2d  &        97.0   &         94.8     &\textbf{94.9}  \\ 
     & PCA   2d  &        69.9   &         65.5     &        49.8 \\\bottomrule
\end{tabular}
\caption{F1 macro score results on \emph{member} datasets using logistic regression with RE user representations on strongly (leave-one-out cross validation) and weakly (3- and 1-shot) supervised scenarios.}
\label{tab:f1_results_weakly}
\end{table}

\section{Experiment \#2: Weakly Supervised Scenario}

\paragraph{\textbf{Experiment Settings.}} We now experiment with a more challenging, weakly supervised scenario, where the models are provided with very limited training data in two different settings: (i) one-shot, where only one user is selected for training per class, and (ii) three-shot, a few-shot setting for which only three users are selected for training per class. The remainder of the users are left for the test set. In the interest of focus and brevity, for this setting we only use the RE user representations and Logistic Regression method, which was the best combination in Experiment \#1 above. Furthermore, we also provide results obtained with and without dimensionality reduction techniques.

\paragraph{\textbf{Results.}} Table \ref{tab:f1_results_weakly} shows that 2 dimensional representations derived from t-SNE and UMAP dimension reduction technique get better results than RE without any dimensionality reduction for one-shot and few-shot settings. Results with PCA reduction are substantially worse across evaluation settings and regions. Compressing RE user representations into 2 dimensional representations with UMAP or t-SNE can be a good solution to handle community detection on weakly supervised scenarios as they can highlight communities due to their architecture based on nearest-neighbours. Interestingly, despite being evaluated on few-shot and one-shot settings, these methods obtain scores similar to those of RE on the strongly supervised scenario. Furthermore, results are consistent across the 3 regions, showing that RE representations reach competitive and robust results even in weakly supervised scenarios.

\begin{table*}[ht!]\footnotesize
\centering
\begin{tabular}{@{}ll|rrrrr|rrrrr|rrrrr@{}}
\toprule
\multicolumn{2}{c}{} & \multicolumn{5}{c}{SCT}  & \multicolumn{5}{c}{WAL}  & \multicolumn{5}{c}{NIR}   \\ \midrule
&&Mj&N2V&DW&FA2&RE  &Mj&N2V&DW&FA2&RE &Mj&N2V&DW&FA2&RE\\ \midrule
Supporter & LogReg   
& 6.8  & 23.0 & 24.7 & 21.6 & \textbf{90.8}
& 10.1 & 38.8 & 48.2 & 27.2 & \textbf{95.3}
& 7.8  & 21.2 & 20.9 & 31.1 & \textbf{75.4} \\
& RF    
& 6.8  & 50.8 & 44.1 & 40.3 & \textbf{81.4}
& 10.1 & 50.0 & 59.9 & 56.1 & \textbf{93.8}
& 7.8  & 30.5 & 33.3 & 36.8 & \textbf{65.4} \\\midrule
Sympathizer & LogReg           
& 7.1  &  8.3 & 09.4 & 18.0 & \textbf{63.3}
& 10.6 & 19.8 & 17.6 & 23.3 & \textbf{60.6}
& 7.4  &  6.5 &  6.5 & 21.6 & \textbf{41.2} \\
& RF    
& 7.1  & 22.8 & 20.7 & 25.6 & \textbf{51.7}
& 10.6 & 17.8 & 17.2 & 23.8 & \textbf{60.2}
& 7.4  &  9.3 &  7.7 & 26.3 & \textbf{38.1} \\\midrule
& avg. 
& 7.0  & 26.2 & 24.7 & 26.4 & \underline{\textbf{71.8}}
& 10.4 & 31.6 & 35.7 & 32.6 & \underline{\textbf{77.5}}
& 7.6  & 16.9 & 17.1 & 28.9 & \underline{\textbf{55.0}} \\\bottomrule
\end{tabular}
\caption{F1-score results on SCT, WAL and NIR Supporter and Sympathizer users datasets. Mj refers to majority label classifier used as baseline. Algorithms used to generate the user representations: N2V (Node2vec), DW (DeepWalk), FA2 (ForceAtlas2) RE (Relational Embeddings). Underlined values represent statistically significant ($p$ $<$ $0.05$) highest average values.}
\label{tab:f1_results_aug}
\end{table*}

\section{Experiment \#3: Realistic Scenario}

\paragraph{\textbf{Experiment Settings.}} We define a more realistic, challenging scenario, in which we test the ability of the models to predict the political leaning of less engaged users, namely, of supporters and sympathizers. This assessment can offer insights into the model's applicability in a real-world context, where individuals may not be directly affiliated with political parties and have different levels of attachment. In order to do this, we use \emph{members} for training and \emph{supporters} and \emph{sympathizers} for testing. We break down the performance of the models for each of the groups --supporters and sympathizers-- to evaluate the impact of the level of political engagement of users to infer political leaning. For these experiments we use the two best overall classifiers in the strongly supervised scenario: Logistic Regression and Random Forest.

\paragraph{\textbf{Results.}} Table \ref{tab:f1_results_aug} shows that the performance for this scenario is considerably lower. Overall, supporter users get better results than sympathizer users, showing that models suffer more when trying to learn political leaning of users that do not engage so much with political parties (which is only natural, in a way). N2V and DW fail to produce satisfactory results for this realistic scenario. We hypothesize that random walks may generate noisy or irrelevant paths that can negatively affect the quality of the embeddings. The FA2 method also fails to infer political leaning, showing that the use of a two-dimensional vector space for the approximation-repulsion process is insufficient to embed complex socio-political information. In any case, as it has been the case for previous experiments, the pair-based RE user representations are significantly better for every evaluation setting across every region. All the methods without exception show higher results on WAL datasets (4 classes) than on SCT and NIR datasets (5 classes). It seems that including one class more in a multiclass approach to political leaning inference has negative influence to classify users which are less engaged in politics (supporters and sympathizers).

\section{Discussion}

In this section we discuss the analysis of political leaning in light of the reported results. We also consider different ways of making our results explainable and provide an error analysis to identify possible weaknesses of our approach.

\paragraph{\textbf{Explainability.}} 
In order to better understand and explain the effectiveness of the different user representation techniques, we visualize RE, N2V and DW user representations for the three regions, SCT, WAL and NIR by performing t-SNE dimensionality reduction into 2 dimensions.

\begin{figure}[ht!]
\centering
    \includegraphics[width=0.33\linewidth]{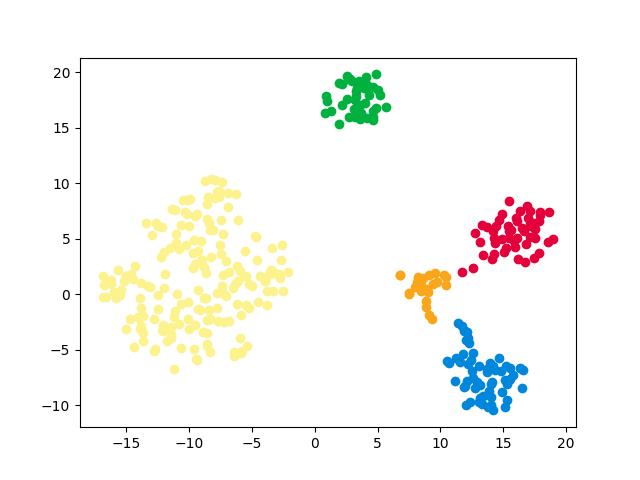} \hfil 
    \includegraphics[width=0.33\linewidth]{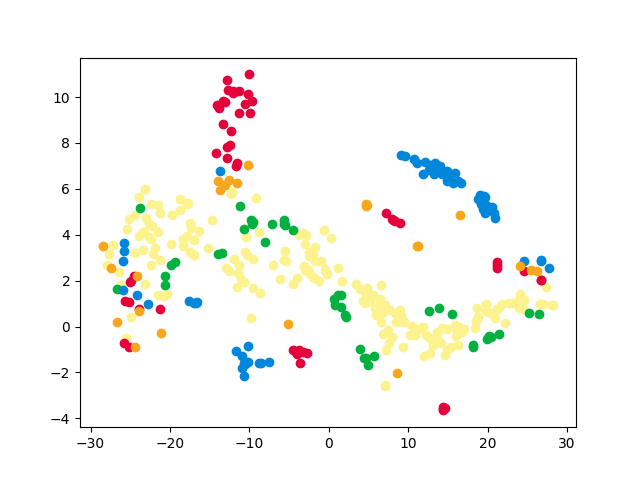}\hfil 
    \includegraphics[width=0.33\linewidth]{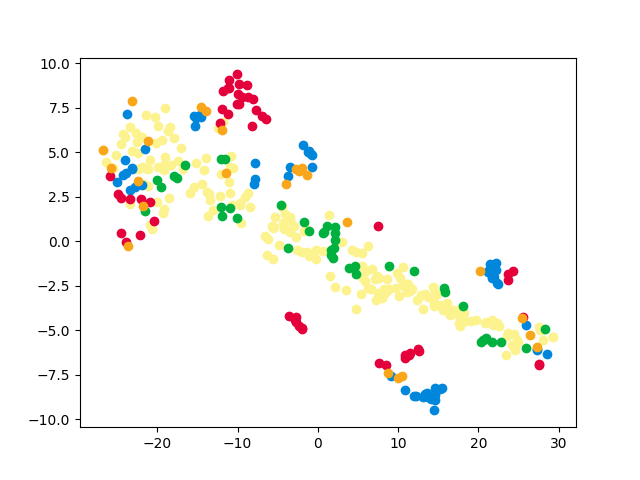}
    \caption{\footnotesize Visualization of t-SNE 2 dimension reduction of Relational Embeddings (left), node2vec (center) and Deep Walk (right) representations for SCT Member users.}
    \label{fig:emb_vis_sct} \hfil
\end{figure}

\begin{figure}[ht!]
\centering
    \includegraphics[width=0.33\linewidth]{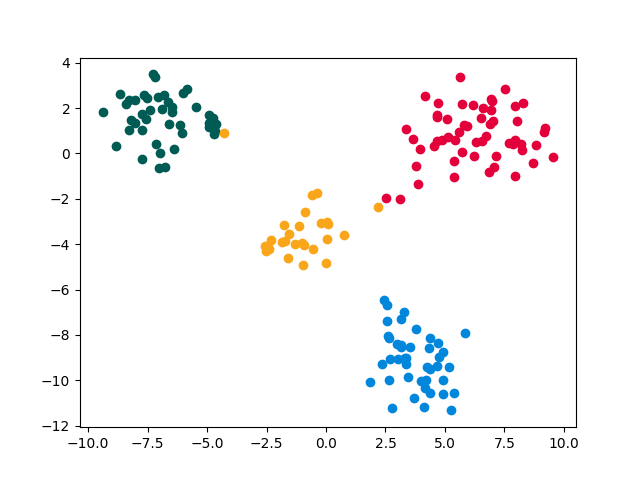} \hfil 
    \includegraphics[width=0.33\linewidth]{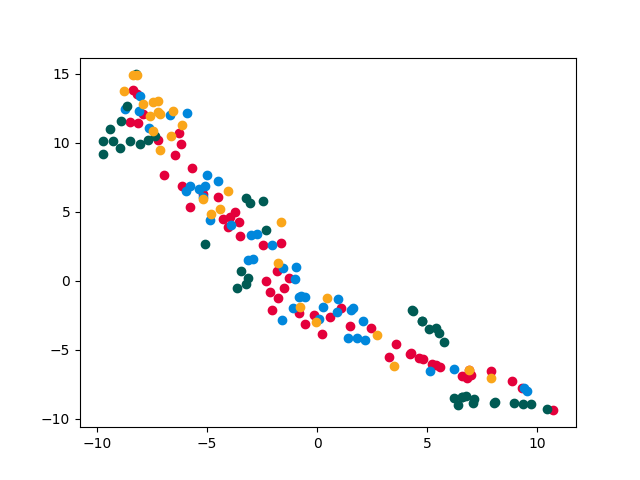}\hfil 
    \includegraphics[width=0.33\linewidth]{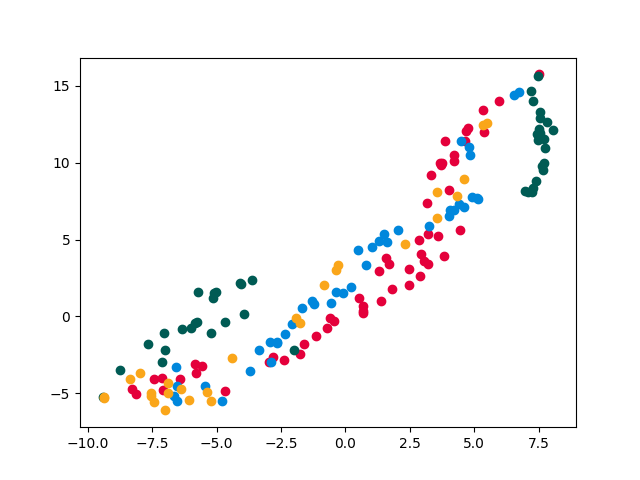}
    \caption{\footnotesize Visualization of t-SNE 2 dimension reduction of Relational Embeddings (left), node2vec (center) and Deep Walk (right) representations for WAL Member users.}
    \label{fig:emb_vis_wal} \hfil
\end{figure}

\begin{figure}[ht!]
\centering
    \includegraphics[width=0.33\linewidth]{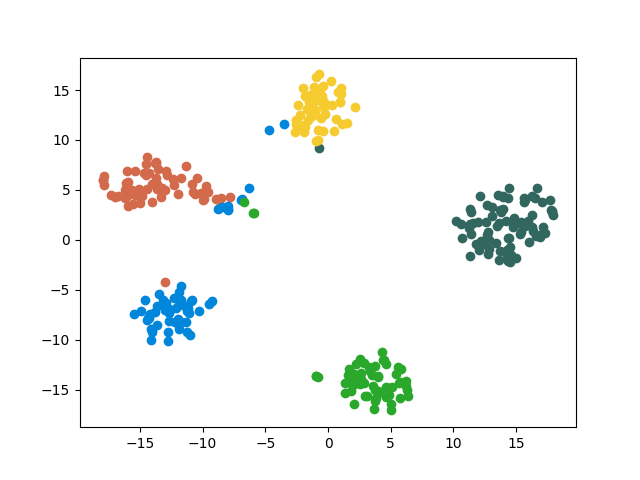} \hfil 
    \includegraphics[width=0.33\linewidth]{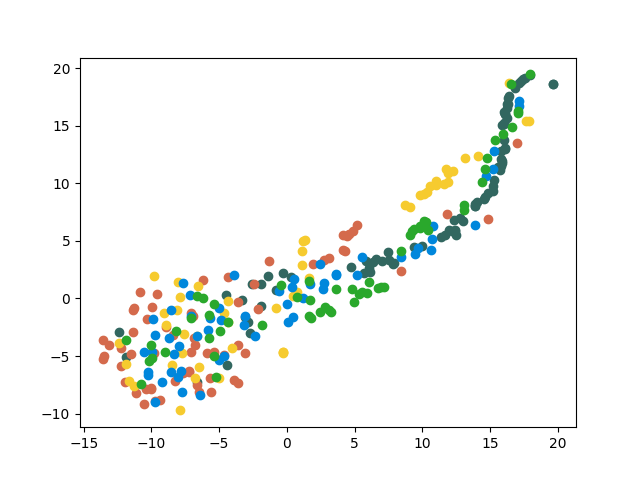}\hfil 
    \includegraphics[width=0.33\linewidth]{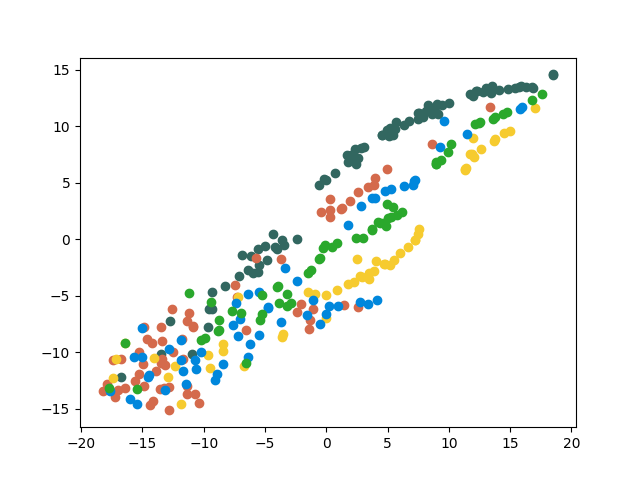}
    \caption{\footnotesize Visualization of t-SNE 2 dimension reduction of Relational Embeddings (left), node2vec (center) and Deep Walk (right) representations for NIR Member users.}
    \label{fig:emb_vis_nir} \hfil
\end{figure}

The first noticeable point looking at Figures \ref{fig:emb_vis_sct}, \ref{fig:emb_vis_wal} and \ref{fig:emb_vis_nir} is that, in contrast to N2V and DW, the visualizations obtained from the RE representations are clearly able to discriminate the multiparty political communities represented by the member users for each of the countries. In fact, the clear communities obtained in the visualization of the RE user representations is arguably in accordance with them outperforming other methods in the experimental evaluations reported in Tables \ref{tab:f1_results_sup} and \ref{tab:f1_results_aug}.

Taking a closer look at the SCT visualization (left plot in Figure \ref{fig:emb_vis_sct}), we can see parties represented by clearly distinguishable communities. Thus, SNP (\textcolor{snp}{$\bullet$}) takes a big part of the figure, mainly isolated from the others. Unionist parties are forming their own cluster at the right side of the chart, separated from the other two parties. Inside the unionist community, it can be seen that SLD (\textcolor{lib}{$\bullet$}) acts like a link between SCU (\textcolor{con}{$\bullet$}) and SL (\textcolor{lab}{$\bullet$}), showing its position as a central political actor.  SLD also takes a central role in the which highlights its centrist political outlook. The representation locates SGP (\textcolor{sgp}{$\bullet$}) apart from SNP and the unionists but between SNP and SL, showing a proximity to those.  Furthermore, it can be seen that the pro-independence (SNP and SGP) and unionist parties (SL, SLD and SCU) are represented in different positions, showing a high polarization in the dispute across national identities.

Moving on to Figure \ref{fig:emb_vis_wal} which shows the visualization obtained for WAL, it is possible to note that WLD (\textcolor{lib}{$\bullet$}) is situated in the center of the political spectrum, which is interesting as in reality they are considered to be positioned in the political center. The other 3 parties are surrounding WLD, but WL (\textcolor{lab}{$\bullet$}) is between PC (\textcolor{pc}{$\bullet$}) and WC (\textcolor{con}{$\bullet$}), showing that the last two are the most opposed poles (left/right or pro-independence/unionist).      

Finally, for NIR we can see in Figure \ref{fig:emb_vis_nir} that political parties are grouped in their own clusters, except for a few instances that may have been incorrectly classified. These few errors seem to align with to those causing slightly lower performance, as shown in Table \ref{tab:f1_results_sup}, of the RE user representations for NIR when compared to SCT and WAL. Looking at the positions of the parties, we can see that DUP (\textcolor{dup}{$\bullet$}) is located next to UUP (\textcolor{con}{$\bullet$}) forming a conservative and unionist pole. Besides, SF (\textcolor{sf}{$\bullet$}) and SDLP (\textcolor{sdlp}{$\bullet$}) define the left-wing and pro-Irish pole. As a centralist political actor APNI (\textcolor{apni}{$\bullet$}) is located between both main groups, but much closer to the conservative-unionist pole forming with them a wider liberal-conservative, right-wing pole at the left of the chart. Summarizing, we believe that REs capture well multiple ideological disparities (left/right or pro-Irish/unionist) among these parties.

The ability of RE user representations to depict distinct communities in all three cases is noteworthy, especially when compared to the inability of N2V and DW. Furthermore, REs are able to locate the communities following a pattern of ideological similarities and disparities. These observations lead us to infer that RE user representations have the potential to embed socio-political information within the generated features.

\paragraph{\textbf{Data quantity.}} 
Relating the results with the number of the interactions collected in the datasets, we can see that there are some similarities among the user representation methods. The RE user representation method has better results for strongly supervised approach at SCT (19M RTs) and WAL (21M RTs) comparing to a small drop for NIR (4M RTs) which contains considerably less interactions (Table \ref{tab:data_final}). The same occurs in the weakly supervised scenario, in which SCT and WAL obtain better evaluation results in comparison to NIR. This not only occurs with REs, but also with other user representation methods such as N2V and DW. The take out message is seems to be that the larger the number of interactions the better the results. The results for the realistic scenario reported in Table \ref{tab:f1_results_weakly} further confirm this trend. Thus, the larger the timelines and the amount of users from which to extract the retweets, the better representations we get for all the user representation methods.

\begin{figure}[ht!]
\centering
    \includegraphics[width=0.49\linewidth]{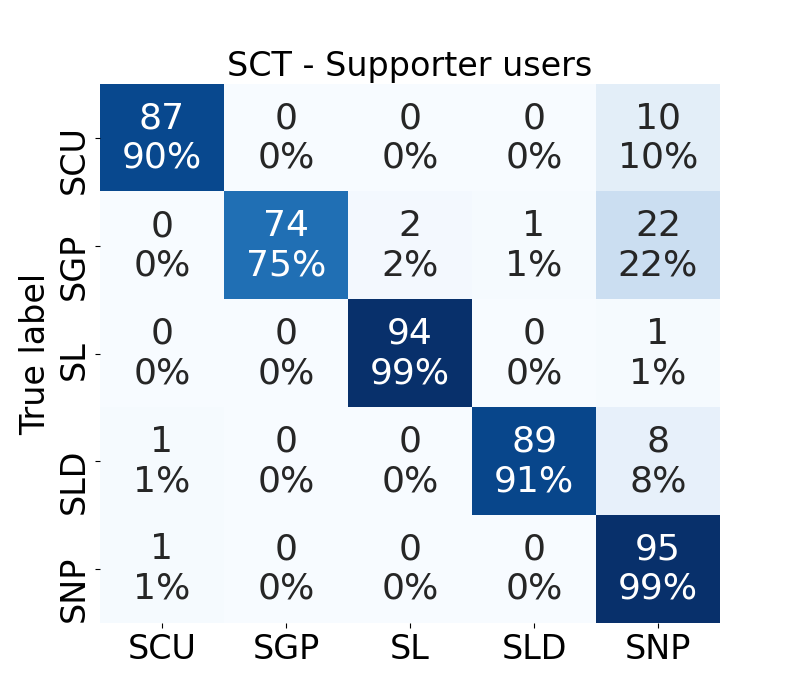}\hfil 
    \includegraphics[width=0.49\linewidth]{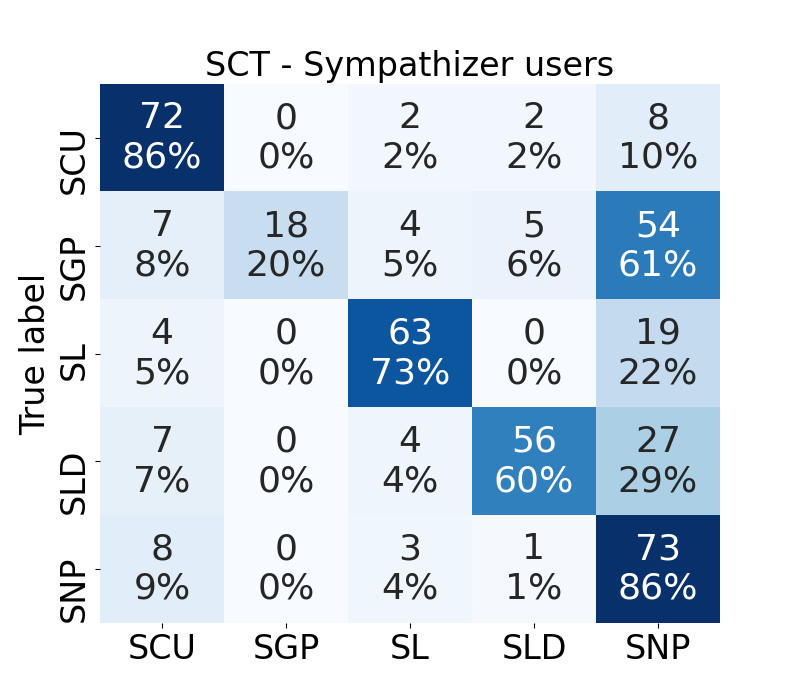}
    \caption{\footnotesize Confusion matrices of SCT Supporter (left) and Sympathizer (right) users of Logistic Regression trained with RE user representations.}
    \label{fig:hm_sct} \hfil
\end{figure}

\begin{figure}[ht!]
\centering
    \includegraphics[width=0.49\linewidth]{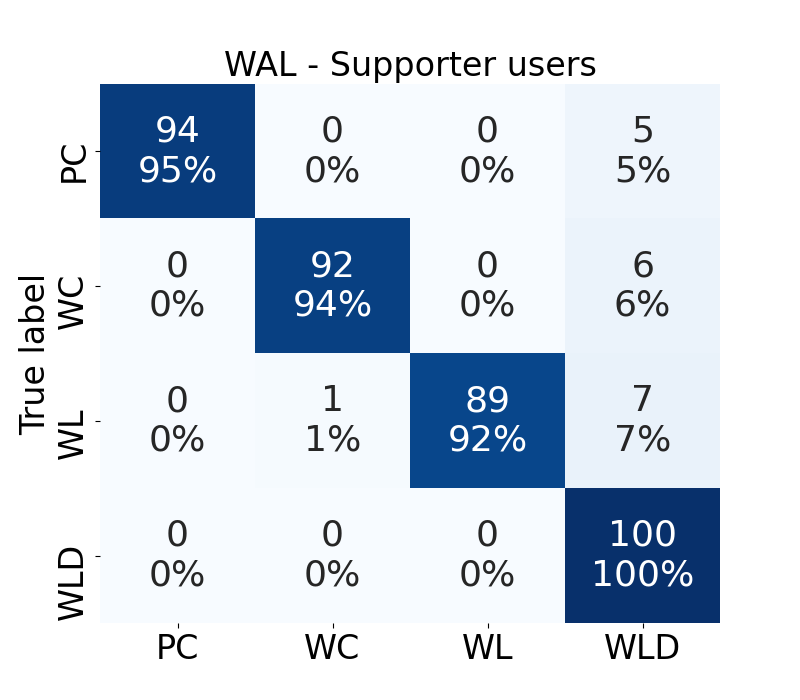}\hfil 
    \includegraphics[width=0.49\linewidth]{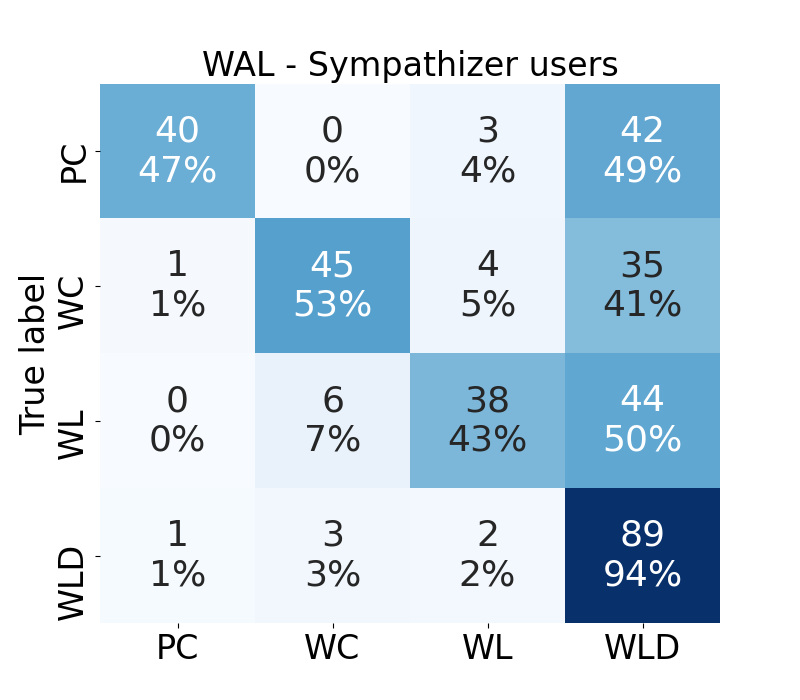}
    \caption{\footnotesize Confusion matrices of WAL Supporter (left) and Sympathizer (right) users of Logistic Regression trained with RE user representations.}
    \label{fig:hm_wal} \hfil
\end{figure}

\begin{figure}[ht!]
\centering
    \includegraphics[width=0.49\linewidth]{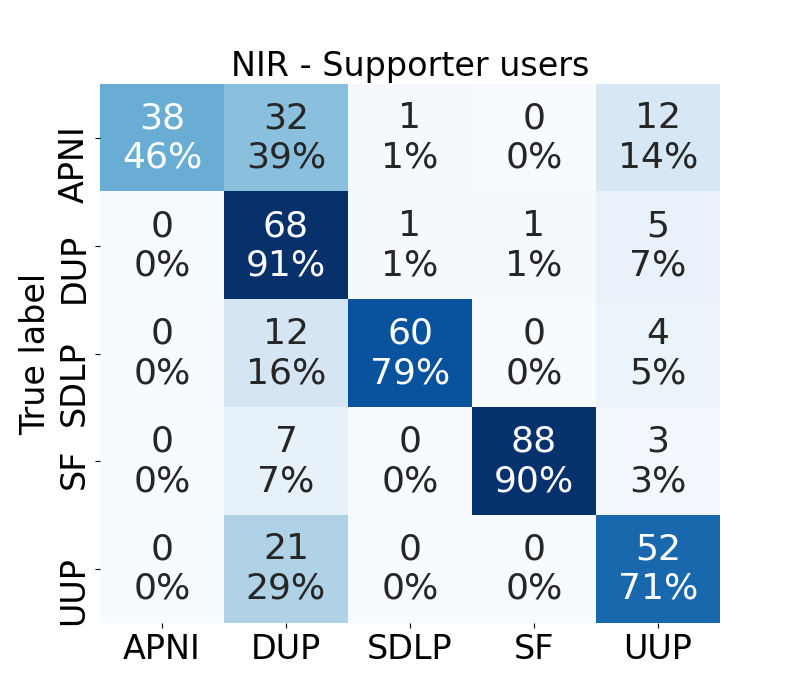}\hfil 
    \includegraphics[width=0.49\linewidth]{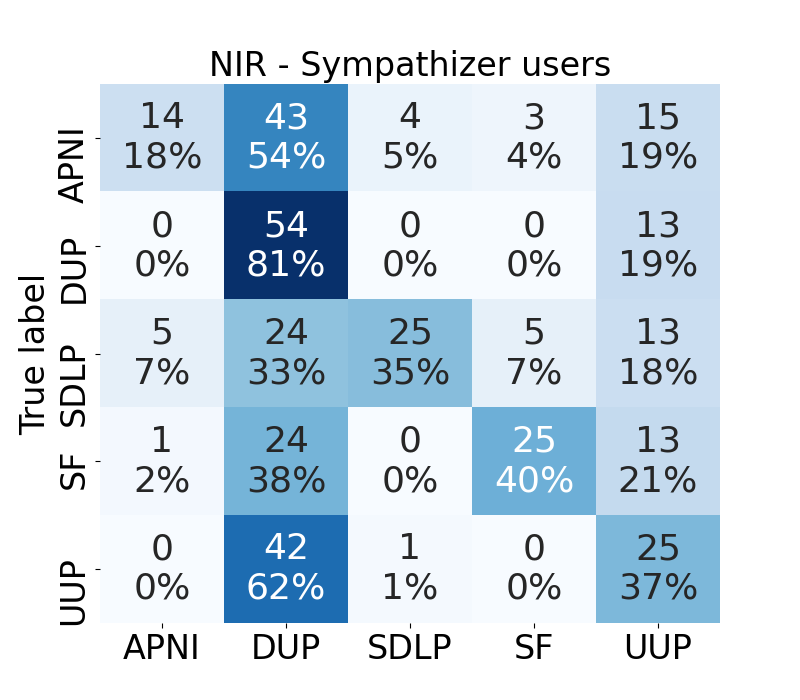}
    \caption{\footnotesize Confusion matrices of NIR Supporter (left) and Sympathizer (right) users of Logistic Regression trained with RE user representations.}
    \label{fig:hm_nir} \hfil
\end{figure}

\paragraph{\textbf{Error analysis.}}
Considering the almost perfect scores obtained by REs on the strongly and weakly supervised scenarios, our error analysis will focus only on the realistic scenario, which consists of users less engaged politically, namely, supporters and sympathizers. The confusion matrices presented in Figures \ref{fig:hm_sct}, \ref{fig:hm_wal} and \ref{fig:hm_nir} report the main errors performed by the classifiers based on REs on this particular scenario.

With respect to SCT, it can be observed that most misclassified instances correspond to models predicting SNP instead of the correct label. Thus, for supporter users (Figure \ref{fig:hm_sct} left), 22\% of the errors in predicting SGP users (75\% acc.) are wrongly predicted as SNP.  This is even more pronounced for sympathizer users (Figure \ref{fig:hm_sct} right) given that the classifiers performs substantially worse in this evaluation setting. Thus, 61\% of the errors in classifying SGP correspond to the model predicting SNP instead. We hypothesize that this may be explained by the fact that SNP and SGP have certain ideological similarities and have been in a cooperation agreement since 2021. 

The model trained for WAL, compared to the SCT case, has a lower error rate when predicting supporter's political leaning (Figure \ref{fig:hm_wal} left). There, the few errors correspond to predicting WLD instead of the correct classes. If we look at the sympathizers errors (Figure \ref{fig:hm_wal} right), they substantially amplify the pattern seen for in the supporters setting. We believe that the source of errors may be caused by the central role played by WLD in Wales's politics and by the plurality in policies across the political spectrum.

With respect to NIR, we can see that the model has more problems discriminating between the different political options. If we look at the confusion matrix for supporters (Figure \ref{fig:hm_nir} left), DUP gets 32\% of UUP and 39\% of APNI instances. Moreover, 12\% of the APNI errors correspond to UUP, which shows that the model is not able to detect well APNI users. This might be due to the centralist and liberal profile of APNI, which makes it difficult to discriminate from other political options. The same pattern can be observed for the sympathizer users (Figure \ref{fig:hm_nir} right) although amplified by the larger number of classification errors. It is particularly interesting the difficulties of the model in distinguishing UUP and APNI from DUP, which seems to indicate that the DUP is seen as the main reference of the right unionist space.

The confusion matrices in the realistic scenario show that the best model suffers to clearly classify multi-party political leaning especially for users less engaged politically (sympathizers). It is noticeable the large amount of errors when trying to discriminate UUP and APNI from DUP, which is indicative perhaps of the dominance by DUP of the right wing agenda. In the case of WAL, the main source errors seem to be due to the centralist position of WLD with respect to other political options. Finally, in SCT we can see the phenomenon of a smaller party (SGP) cooperating with a larger one (SNP) and getting assimilated as a result of this collaboration. Summarizing, our political party-based outlook may capture sociopolitical information since errors commonly occur among ideologically adjacent classes and increase when targeting less politically engaged users. In any case, we confirm the necessity of addressing political leaning as a multipolar classification task which, despite being more difficult, would provide a more representative analysis of the social reality.

\section{Conclusion and Future Work} 

In this work we look at the ability to infer the political leaning of social media users across multiple regions with multi-party systems, a challenging scenario that, to the best of our knowledge, has not been studied before. In order to do this, we collect a dataset spanning three UK regions, where users with different levels of political engagement (members, supporters, sympathizers) are labelled by the political party they align with. By conducting a set of experiments with these three datasets, we find that a model leveraging user interactions based on Relational Embeddings, in combination with a Logistic Regression classifier, achieves the best results. Unlike the other methods, REs use real user interactions without generating any artificial user connections \citep{fernandez2022relational}. Experimental results are consistent across the three regions and different political engagement, demonstrating its robustness. However, experiments also show that predictions get particularly more challenging as the level of engagement of users decreases. Parallel to other social behaviors, less attachment means fewer performative actions that may define the political preferences of an individual, becoming more difficult to infer. Finally, visualizations and error analysis evidenced that REs are capable of capturing socio-political information.

There are other avenues for future research which were not within the scope of this work but would be worth exploring. For example, the use of other features beyond interactions could be potentially useful, as it could be the use of textual data to improve classification results for sympathizers.  This would also enable the extensibility of the model to other social media, where retweet interactions do not exist. Moreover, in order to develop a richer political analysis, we are eager to apply the data extraction and user representation techniques into other tasks such as propaganda and disinformation detection.

\bibliography{bibliography}

\end{document}